\documentclass[useAMS,usenatbib]{mn2e}

\usepackage{natbib}

\usepackage{latexsym,graphicx}
\usepackage{color}
\usepackage{fixltx2e}
\usepackage{verbatim} \usepackage{float}
\usepackage{amsmath,amssymb}
\usepackage{times}

\usepackage{setspace}
\usepackage{rotating}
\usepackage{pdflscape}
\usepackage{subfig}

\usepackage[squaren, Gray, cdot]{SIunits}

\def\apgt{\ {\raise-.5ex\hbox{$\buildrel>\over\sim$}}\ }
\def\aplt{\ {\raise-.5ex\hbox{$\buildrel<\over\sim$}}\ }

\let\oldhat\hat
\renewcommand{\hat}[1]{\oldhat{\mathbf{#1}}}

\newcommand{\mpo}{\textcolor{black}}

\title[Ring nebulae around runaway Wolf-Rayet stars]{On the ring nebulae around runaway Wolf-Rayet stars}

\author[D. M.-A.~Meyer et al.]
       {D. M.-A.~Meyer\thanks{E-mail: dmameyer.astro@gmail.com}$^{1}$, L.~M.~Oskinova$^{1,2}$, M.~Pohl$^{1,3}$ and M.~Petrov$^{4}$ \\
       $^{1}$ Universit\"at Potsdam, Institut f\"ur Physik und Astronomie, Karl-Liebknecht-Strasse 24/25, 14476 Potsdam, Germany\\
       $^{2}$ \textcolor{black}{Department of Astronomy, Kazan Federal University, Kremlevskaya Str 18, Kazan, Russia} \\
       $^{3}$ DESY, Platanenallee 6, 15738 Zeuthen, Germany \\        
       $^{4}$ Max Planck Computing and Data Facility (MPCDF), Gießenbachstrasse 2, D-85748 Garching, Germany\\ 
       }  

\begin{document}

\date{Received; accepted}

\maketitle
   
\label{firstpage}

\begin{abstract} 
Wolf-Rayet stars are advanced evolutionary stages of massive 
stars. Despite their \textcolor{black}{large} mass-loss rates and \textcolor{black}{high} wind velocities, 
none of them display a bow shock, although a fraction of them are classified 
as runaway. 
Our \textcolor{black}{2.5-D} numerical simulations of circumstellar matter  around a 
$60\, \rm M_{\odot}$ runaway star show that the fast Wolf-Rayet stellar wind is released 
into a \mpo{wind-blown cavity filled with} various shocks and discontinuities generated 
throughout the precedent evolutionary phases.  
The resulting fast-wind slow-wind interaction leads to the formation 
of spherical shells of swept-up dusty material 
\textcolor{black}{similar to those observed \textcolor{black}{in} near-infrared 
$24\, \rm \mu \rm m$ with \textit{Spitzer}}, and \textcolor{black}{which} appear to 
be co-moving with the runaway massive stars, regardless of their proper motion 
and/or \mpo{the properties of the} local ambient medium. 
We interpret bright infrared rings around runaway Wolf-Rayet stars in the Galactic 
plane, like WR138a, as \mpo{indication of} \textcolor{black}{their very} high initial masses 
\textcolor{black}{and a complex evolutionary history}. 
Stellar-wind bow shocks become faint as stars run in diluted media, therefore,   
our results explain the absence of detected bow shocks around Galactic 
Wolf-Rayet stars such as the high-latitude, very fast-moving  
objects WR71, WR124 and WR148. 
Our results show that the absence of a bow shock is consistent with a runaway 
nature of some Wolf-Rayet stars.
This \textcolor{black}{questions} the in-situ star formation scenario 
of high-latitude Wolf-Rayet stars in favor of dynamical ejection from 
birth sites in the Galactic plane. 
\end{abstract}



\section{Introduction}
\label{sect:introduction}

The evolution of massive stars is poorly understood. It is characterised by 
the release of forceful winds whose intensity \textcolor{black}{increases} as the stars evolve. 
\textcolor{black}{Strong} wind interact with the local interstellar medium (ISM), 
leading to the formation of circumstellar wind bubbles structured by 
several shocks and discontinuities~\citep{weaver_apj_218_1977}. 
These pc-scale shells reflect both wind and ISM properties. 
They constitute the imprint of the past stellar evolution of 
massive stars onto their ambient medium~\citep{meyer_2014a}.
Particularly, high-mass stars are expected to evolve through the 
so-called Wolf-Rayet \textcolor{black}{(WR)} phase. 
Despite the growing consensus on Galactic WR stars as the 
last pre-supernova evolutionary phase of $\ge20\, \rm M_{\odot}$ stars, the 
precise evolutionary channels leading to such stellar objects 
remain uncertain~\citep{crowther_araa_45_2007}. This stage is characterised 
by very strong winds enriched in C, N or O that are blown from the stellar 
surface with velocities up to $ 3000$-$5000\, \rm km\, \rm s^{-1}$ 
and large mass-loss rates reaching $ 10^{-5}\, \rm M_{\odot}\, 
\rm yr^{-1}$~\citep{Hamann2006, bestenlehner_aa_570_2014, Sander2012}. 

WR stars often live in binary systems or have a binary evolutionary history.  
These \textcolor{black}{evolved} stars have most likely travelled \textcolor{black}{away} 
from their birth place in the Galactic plane, 
either after many-body gravitational interaction within stellar groups or 
under the influence of the kick given by the shock wave of a defunct binary companion 
which ended its life as a supernova explosion~\citep{Moffat1979, dincel_mnras_448_2015}. 
However, only a handful of runaway  WR stars are known in the Galaxy. 
This could be because the measurement of \mpo{the radial velocity is complicated for}
WR optical spectra \textcolor{black}{dominated by broad emission lines}. Therefore, often the position 
of a WR star in the high-latitude region of the Milky Way serves 
as the primary indication of its runaway nature. 
The three most-studied high-galactic latitude WR stars 
are the binary WR\,148  \citep[WN8+B3IV, ][]{Munoz2017}, WR\,124 together with 
its suspected relativistic companion~\citep[WN8h, ][and references therein]{Toala2018},
and WR\,71 \citep[WN6, ][]{Moffat1998}. 

\textcolor{black}{The traditional} method to discover runaway stars \textcolor{black}{involves} observations 
of bow-shocks. Indeed, one could expect that fast-moving WR stars are 
prone to produce bow shock nebulae~\citep{wilkin_459_apj_1996} under the 
conjugated influence of both their fast, dense stellar winds and their rapid 
bulk motion. This picture has been supported by several numerical hydrodynamical 
investigations of WR wind-ISM interaction predicting either the formation of 
unstable bow shocks~\citep{vanmarle_aa_444_2005} or unstable ring-like shells~\citep{brighenti_mnras_277_1995}.  
However, no observational evidence of a bow shock around a WR star 
has been reported so far. 
\textcolor{black}{On the other hand}, WR stars are commonly surrounded by nebulae with a 
variety of morphologies, including ring-nebulae~\citep{Miller-Chu1993, Marston1997, 
toala_aa_587_2015d} that have been observed in optical 
H$\alpha$ and [O\,{\sc iii}] lines as well as in 
the mid-infrared~\textcolor{black}{\citep{barniske_aa_486_2008}}. Recently, 
WR nebulae have been observed in the mm 
range \citep{fenech_617_aa_2018}. \citet{Graf2012} pointed  out the prevalence 
of nebula around WR stars that only recently entered this 
evolutionary stage.

\textcolor{black}{Interestingly, even fast moving WR stars (with velocities
up to $v_{\star}\simeq200\, \rm km\, \rm s^{-1}$ relative to the ISM) 
are located at the center of compact spherical shells~\citep{2010MNRAS.405.1047G}. 
} 
The question is therefore why fast-moving WR stars do not produce 
bow shock nebulae at all ? How can their surroundings be shaped as a stable, 
circular shell ? And why are those stars systematically centered onto those 
ring-like shells which appear to be co-moving with the star ? 
%

\begin{figure}
        \centering
        \begin{minipage}[b]{ 0.475\textwidth} 
                \includegraphics[width=1.0\textwidth]{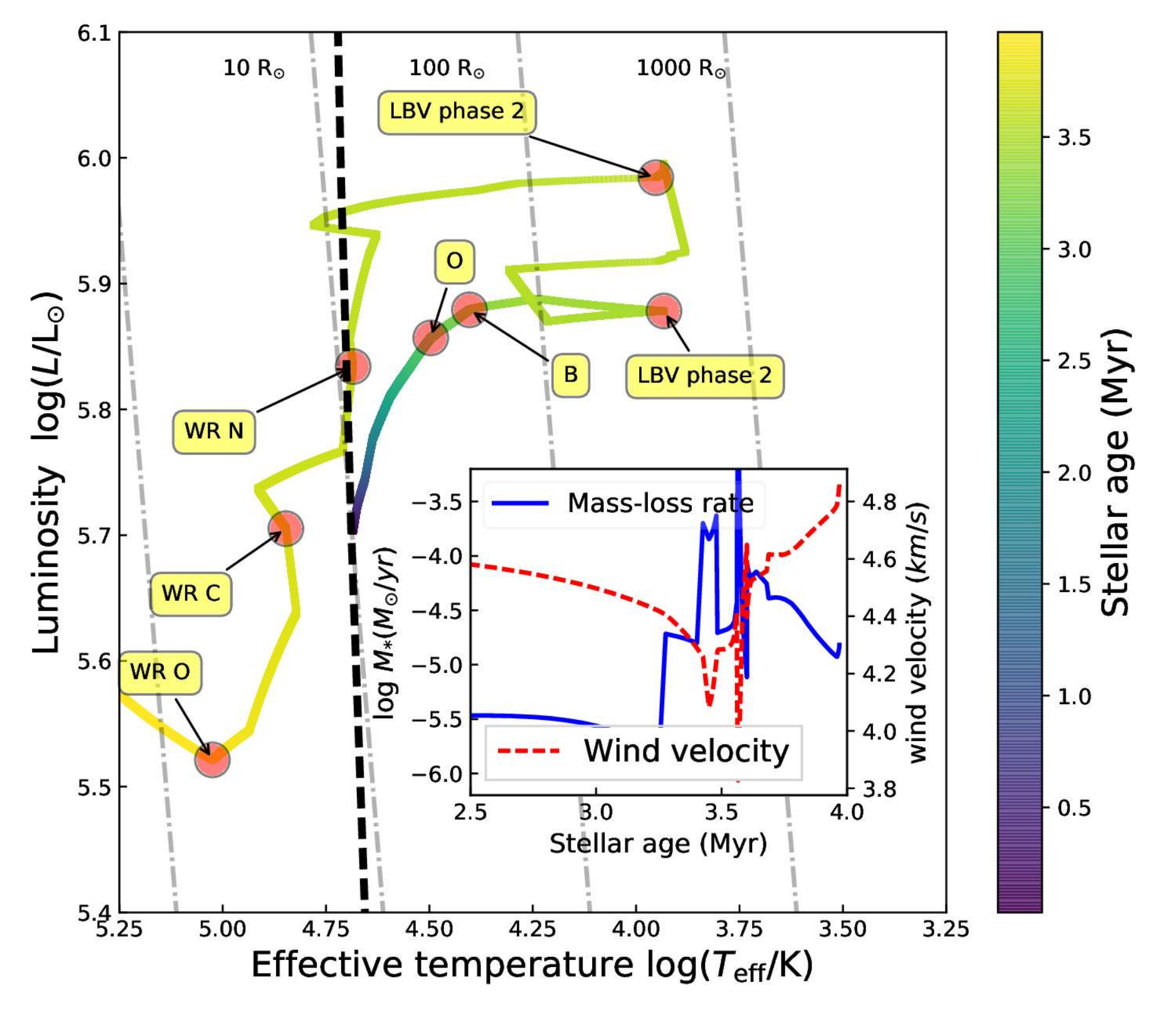}
        \end{minipage}     
        \caption{ 
        \textcolor{black}{
                Evolutionary track of a \mpo{star with an initial mass 
                of $60\,M_{\odot}$} ~\citep{groh_aa564_2014}.
        }
                Inset: mass-loss rate and wind 
                velocity for the post-main-sequence phases. 
                 }      
        \label{fig:hdr}  
\end{figure}

\textcolor{black}{To answer these questions}, we investigate the morphology of 
the circumstellar medium around evolving runaway WR stars by means 
of two-dimensional magneto-hydrodynamical and radiative transfer simulations. 
This study is organized as follows. We review the numerical models 
in Section~\ref{sect:method}, present the ring nebulae models in 
Section~\ref{sect:results}, discuss their significances 
in \textcolor{black}{Section~\ref{section:discussion} and \textcolor{black}{draw conclusions} 
in Section~\ref{section:conclusion}.}

\begin{table}
	\centering
	\caption{
Simulation models. \mpo{Listed are the space velocity of the star, $v_{\star}$, 
and its ambient-medium number density, 
$n_{\rm ISM}$.}  
	}
	\begin{tabular}{lcr}
	\hline
	Model  &  $v_{\star}$ ($\mathrm{km}\, \mathrm{s}^{-1}$)  &  $n_{\rm ISM}$ ($\mathrm{cm}^{-3}$)   \\ 
	\hline    
	Run-v30-n0.79    &  $~~50$     &      $~~~~~~0.79$              \\  
	Run-v100-n0.79   &  $100$      &      $~~~~~~0.79$              \\  
	Run-v200-n0.01   &  $200$      &      $~~~~~~~~~0.01$           \\  
	\hline   
	\end{tabular}
\label{tab:models}\\
\end{table}

\begin{figure}
        \centering
        \begin{minipage}[b]{ 0.475\textwidth} 
                \includegraphics[width=1.0\textwidth]{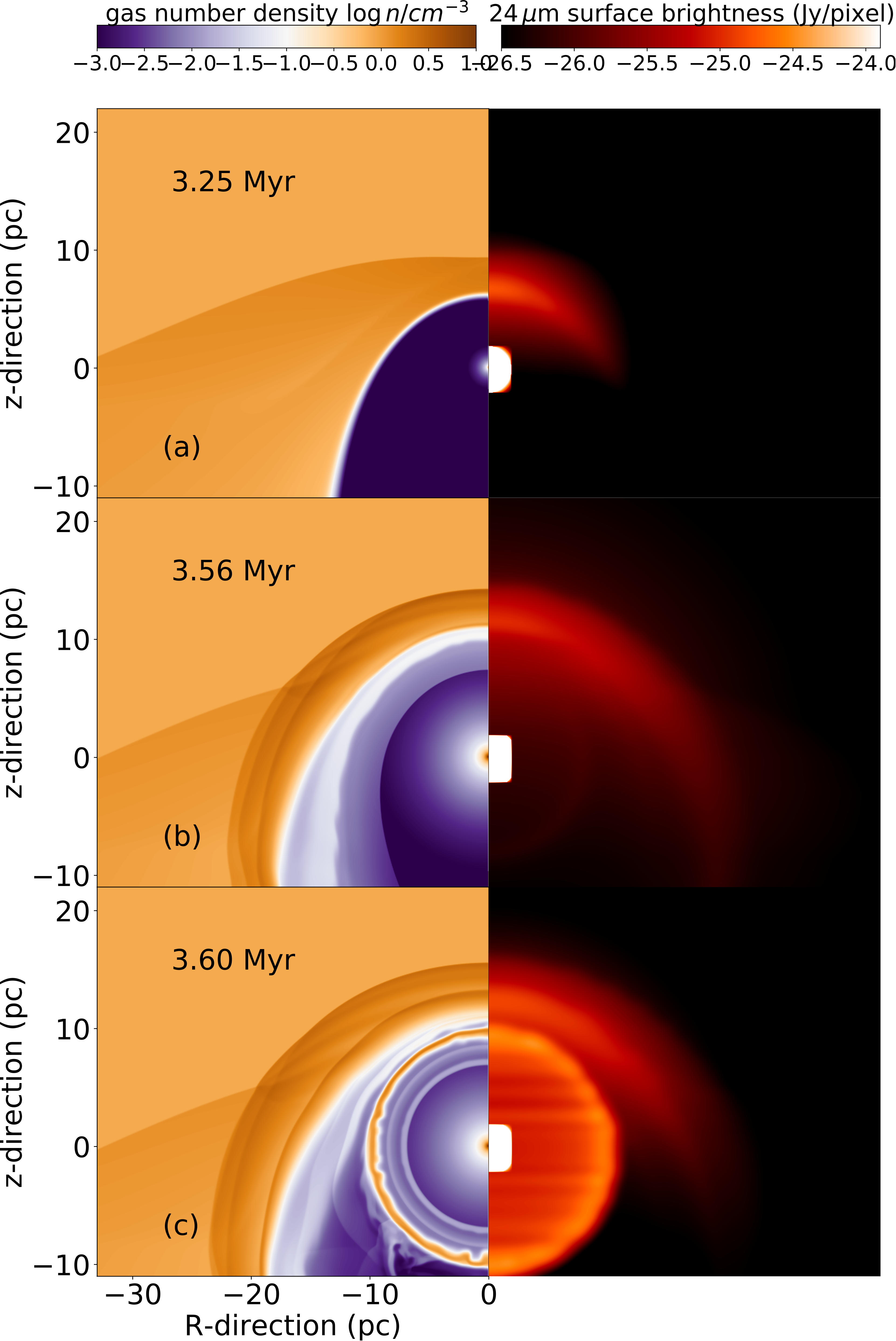}
        \end{minipage}     
        \caption{
                 \textcolor{black}{
                 Density (left) and infrared $24$-$\mathrm{\mu m}$ 
                 emission maps (right) of the circumstellar medium around 
                 a star with initial mass $60\, \rm M_{\odot}$ and evolving 
                 along the evolutionary track shown in Fig.~\ref{fig:hdr}. 
                 It plots the results for the model  Run-v30-n0.79 in which 
                 the star moves with velocity $30\, \rm km\, \rm s^{-1}$ in a 
                 medium of number density $0.79\, \rm cm^{-3}$ which corresponds 
                 to the warm phase of the Galactic plane. 
                 From top to bottom, selected snapshots are displayed, with 
                 the steady-state bow shock of the runaway star's 
                 main-sequence phase (a); the "Napoleon's hat" developing during the B-type 
                 phase of its stellar evolution and making room for the slow LBV materials 
                 to expand inside of it (b); the LBV material swept-up as a mid-infrared 
                 ring nebula by the fast WR wind (c). The spherical ring is brigther 
                 than the bow shock generated by the pre-WR winds. 
                 }
                 }      
        \label{fig:1}  
\end{figure}


\begin{figure}
        \centering
        \begin{minipage}[b]{ 0.485\textwidth} 
                \includegraphics[width=1.0\textwidth]{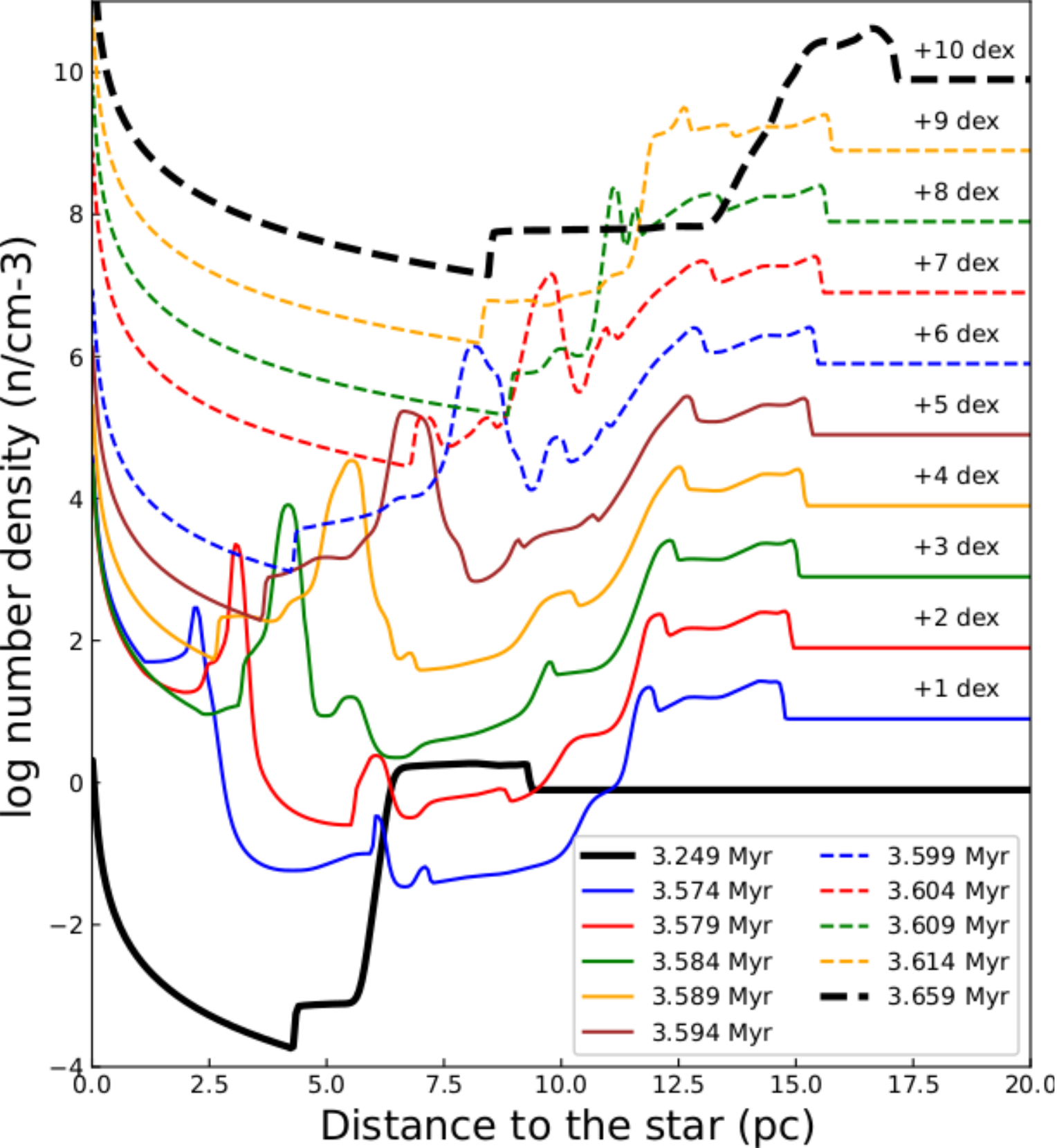}
        \end{minipage}     
        \caption{ 
                 Number density profiles (in $\rm cm^{-3}$) taken along the direction of 
                 motion of the runaway star, at selected times (in $\mathrm{Myr}$) before, during and 
                 after the release of the WR WN shell. Each profile is offset by 1 dex. 
                 }      
        \label{fig:cuts}  
\end{figure}

\section{Numerical simulations}
\label{sect:method}

We perform a series of 2.5D magneto-hydrodynamical (MHD) simulations with the 
{\sc pluto}   code~\citep{migmone_apjs_198_2012}. The simulation have 
been carried out using a cylindrical coordinate system ($\rm R$,$\rm z$) mapped 
with a uniform grid  $[z_{\rm min};z_{\rm max}]\times[0;R_{\rm max}]$ of resolution 
$\Delta=0.08\, \rm pc\, \rm cell^{-1}$ minimum. 
A stellar wind is injected \mpo{in} the computational domain as a half 
sphere of radius $20$ cells centered onto the origin, and a continuous inflow 
is set at the $z=z_{\rm max}$ boundary so that the wind-ISM interaction is modelled 
in the frame of reference of the moving star~\citep{meyer_obs_2016}. 
We consider the ISM material to be at solar metallicity~\citep{asplund_araa_47_2009}. 
Energy loss/gain by optically-thin radiative cooling and heating are taken into account 
using a cooling law for a fully ionized medium~\citep{meyer_2014bb}. Its equilibrium 
temperature is about $ 8000\, \rm  K$, i.e. we assume that the ambient medium is ionized 
by the strong stellar radiation field~\citep{mackey_sept_2014}. 
We assume the stars to move parallely to the local ISM magnetic field of 
strength $7\, \rm \mu G$~\citep{vanmarle_aa_561_2014,meyer_mnras_464_2017}.

We performed a series of 3 numerical simulations. Two ionized 
ISM are considered, \textcolor{black}{with densities} corresponding to either the Galactic plane 
($n_{\rm ISM}=0.79\, \mathrm{cm}^{-3}$) \textcolor{black}{and} to the high Galactic latitudes 
($n_{\rm ISM}=0.01\, \mathrm{cm}^{-3}$). Stars move therein with typical 
space velocity of $30$$-$$100\, \rm km \, \rm s^{-1}$~\citep{renzo_aa_624_2019} 
or $200\, \rm km\, \rm s^{-1}$, respectively. 
We use the stellar\mpo{-evolution} model for a non-rotating $60\, \rm M_{\odot}$ star 
of~\citet{groh_aa564_2014}. 
\textcolor{black}{The track follows the stellar evolution from the zero-age-main-sequence 
through a B supergiant stage, a luminous-blue-variable \textcolor{black}{(LBV)} stage, 
characterized by a few eruptions, and finally the WR phase (Fig.~\ref{fig:hdr}).} 
%
Our simulation \mpo{parameters} are summarised in Table~\ref{tab:models}.

We further post-process our MHD models \mpo{with the code {\sc RADMC-3D}~\citep{dullemond_2012} 
that calculates} the radiative transfer  against dust opacity. 
The dust density field is constructed from the MHD models assuming a standard 
dust-to-gas mass ratio of $1/200$, and the dust temperatures are calculated by 
Monte-Carlo simulations.  
Following~\citet{2019A&A...625A...4G}, the ionized stellar winds and gas hotter 
than $10^{6}\, \rm K$ is considered dust-free. 
Synthetic infrared images are then produced by ray-tracing stellar photons from the 
stellar atmosphere through the stellar surroundings. The star is assumed to be 
a spherical blackbody of effective temperature $T_{\rm eff}$, bolometric 
luminosity $L_{\star}$, and radius $R_{\star}$ interpolated from the stellar 
evolution track of~\citet{groh_aa564_2014}. 
We use opacities for a dust mixture based on silicates crystals~\citep{laor_apj_402_1993}.


\section{Results}
\label{sect:results}

\begin{figure}
        \centering
        \begin{minipage}[b]{ 0.485\textwidth} 
                \includegraphics[width=1.0\textwidth]{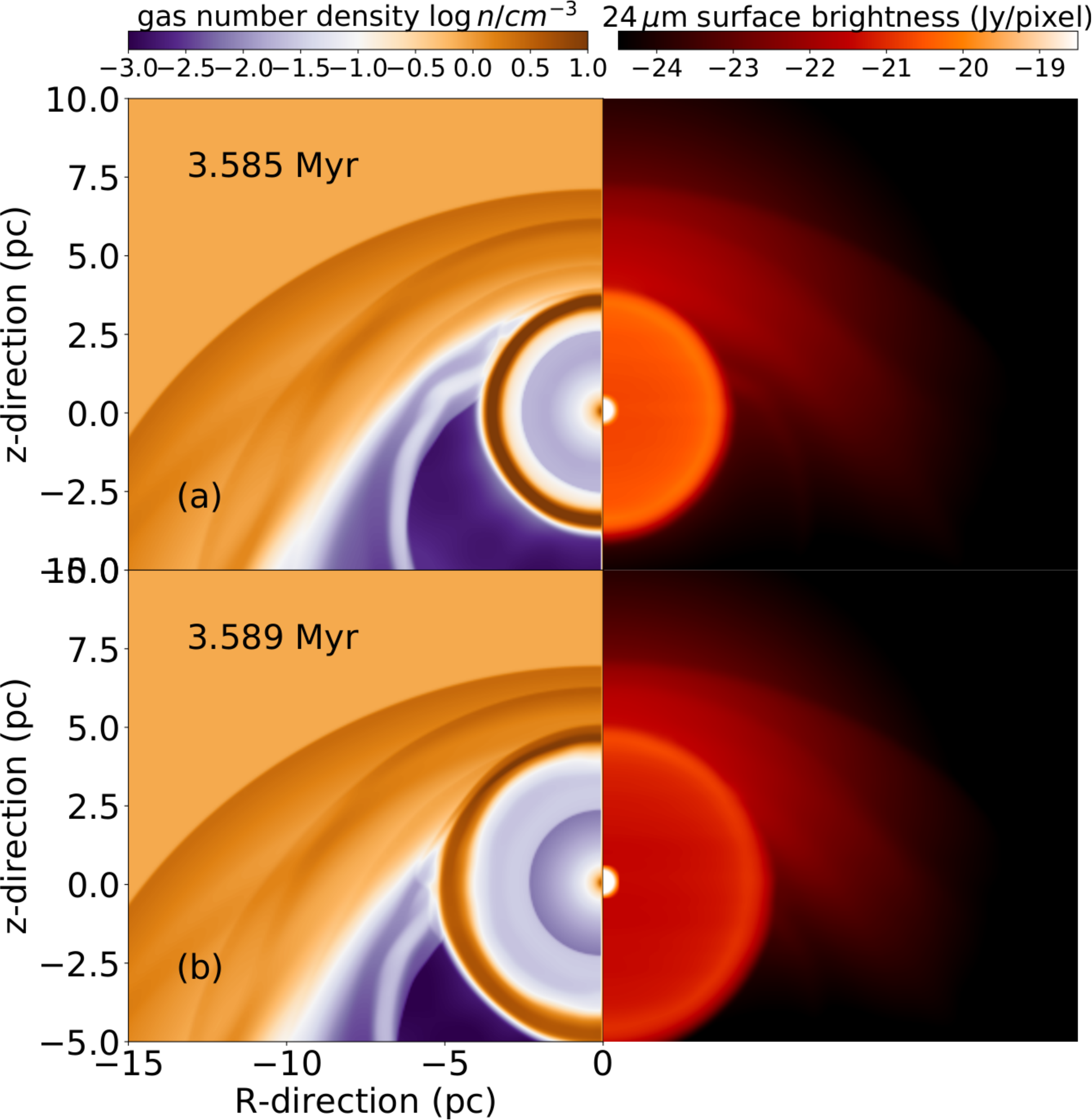}
        \end{minipage}     
        \caption{ 
                 Same as Fig.~\ref{fig:1} for the model Run-v100-n0.79 .  
                 \textcolor{black}{
                 The $60\, \rm M_{\odot}$ star moves with velocity $100\, \rm km\, \rm s^{-1}$ 
                 in a medium of number density $0.79\, \rm cm^{-3}$, corresponding to the warm 
                 phase of the Galactic plane. 
                 From top to bottom, the images show that the fast wind-slow wind 
                 interaction generates a ring nebula around the fast moving WR 
                 star in the Galactic plane (a);  
                 and illustrates how the pre-WR evolution of 
                 the runaway star engenders series of arcs and filaments in the 
                 main-sequence bow shock (b). 
                 This compares well with the H$\alpha$ surroundings of 
                 WR 16~\citep[left panel of Fig.~1 in][]{toala_aa_559_2013}.  
                 }
                 }      
        \label{fig:2}  
\end{figure}

\begin{figure}
        \centering
        \begin{minipage}[b]{ 0.485\textwidth} 
                \includegraphics[width=1.0\textwidth]{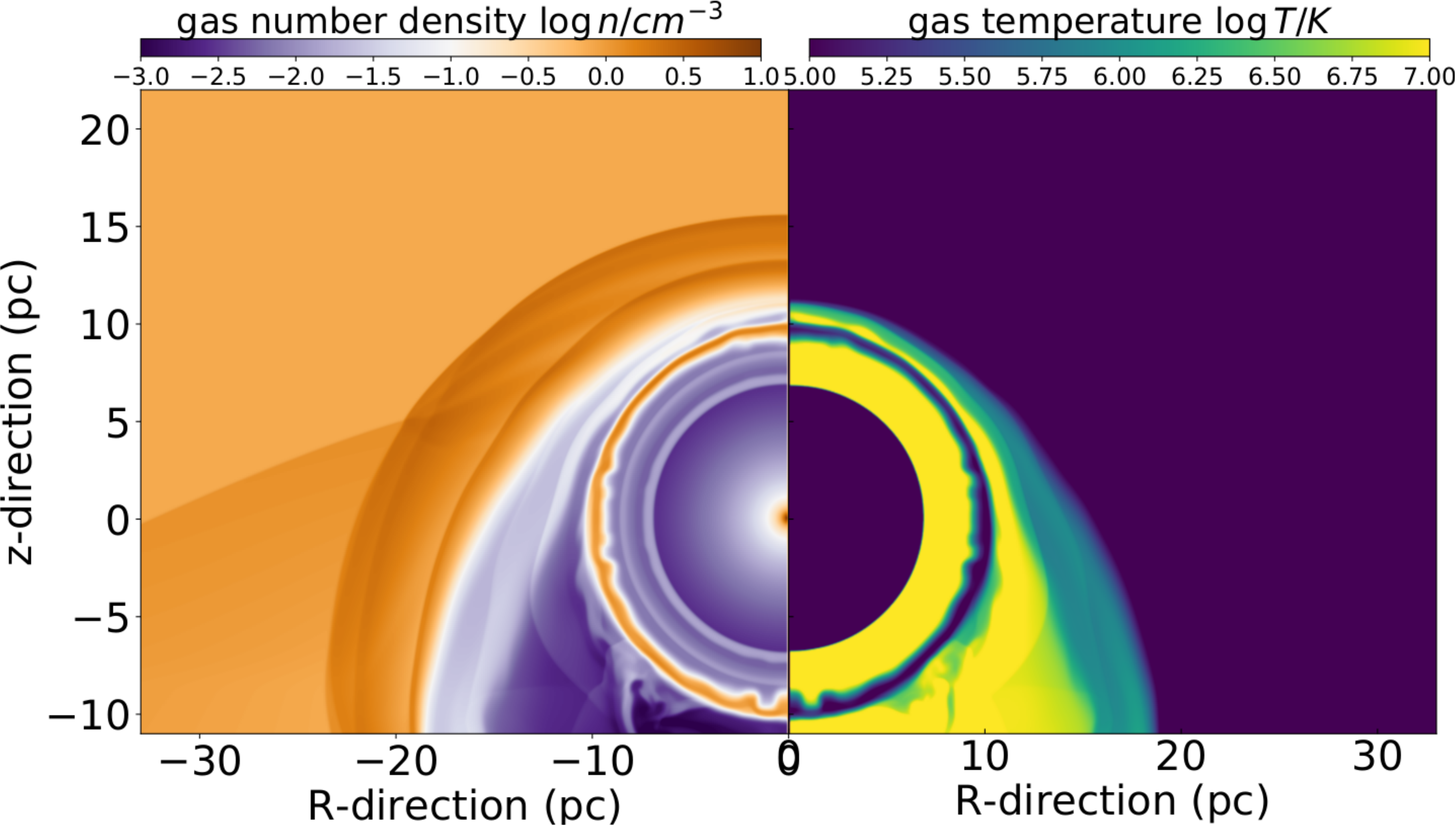}
        \end{minipage}     
        \caption{ 
        \textcolor{black}{
                 Density field (left) and temperature map 
                 of the shocked pre-WR wind and ISM materials (right) 
                 in our model Run-v30-n0.79 at time $3.6\, \rm Myr$. 
                 The right-hand image shows the temperature map of the shock materials 
                 (inner shocked per-WR stellar winds and outer shocked main-sequence 
                 and ISM gas). For clarity, the hot WR freely-expanding wind and the 
                 unperturbed ISM have been substracted from the image. 
                 The swept-up LBV material is colder than its direct surroundings.  
                 Only the ring nebula together with the stellar wind bow shock have 
                 $T\le10^{6}\, \rm K$ and can contain unsublimated dust particles. 
                 The other regions are therefore excluded from the radiative transfer 
                 calculations. 
        }
        }      
        \label{fig:temperature}  
\end{figure}

\begin{figure}
        \centering
        \begin{minipage}[b]{ 0.485\textwidth} 
                \includegraphics[width=1.0\textwidth]{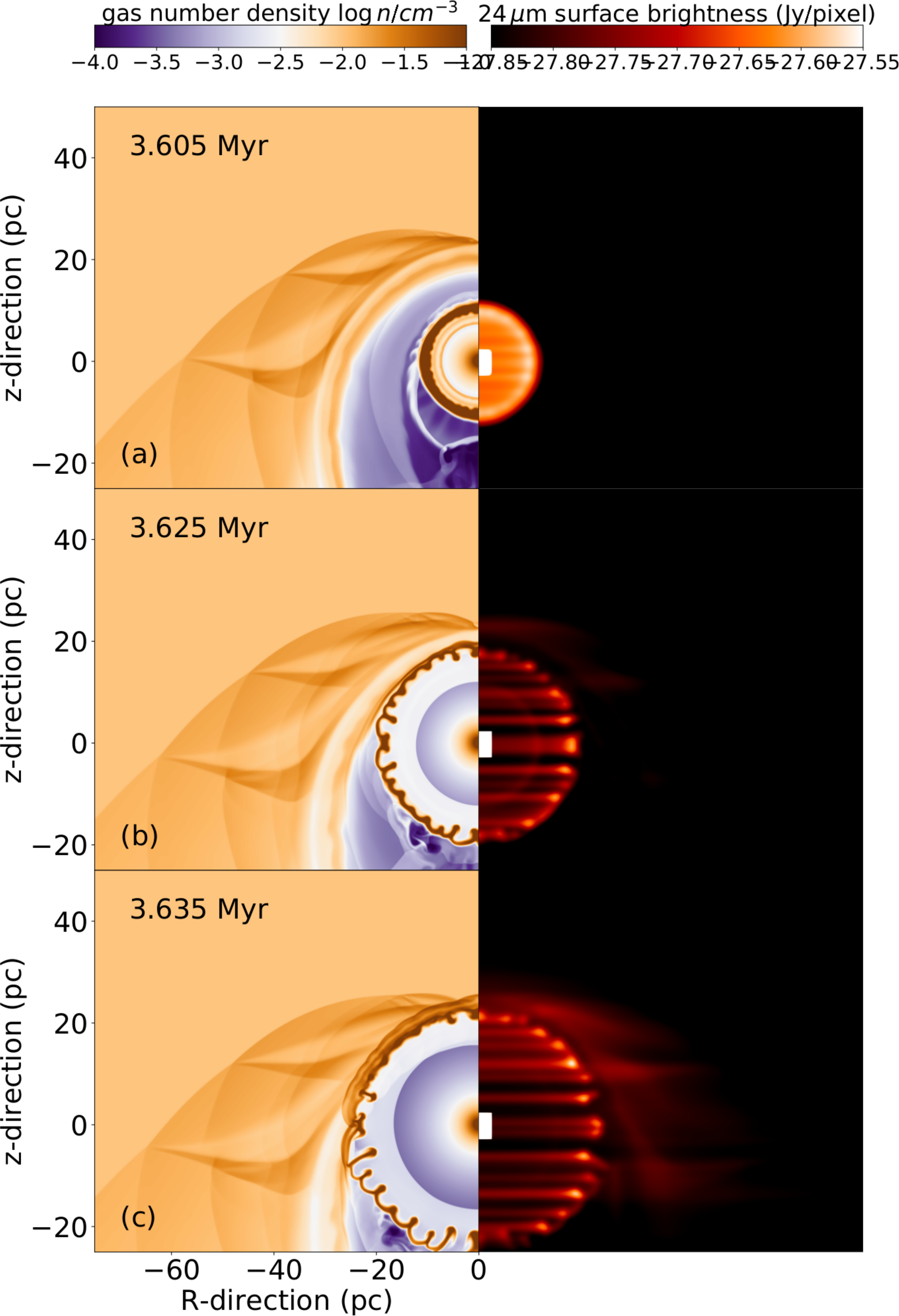}
        \end{minipage}     
        \caption{ 
                 Same as Fig.~\ref{fig:1} but for our model Run-v200-n0.01 .  
                 \textcolor{black}{
                 The star moves with velocity $200\, \rm km\, \rm s^{-1}$ in a 
                 medium of number density $0.01\, \rm cm^{-3}$, corresponding to 
                 the dilute phase of the Galactic plane and/or to the higher latitude 
                 region above the Galactic disc. 
                 As the medium is diluted, the bow shock generated by the runaway star 
                 is huge and its infrared signature is faint, respectively (b). 
                 The region filled by slow LBV material is large, which permits to 
                 the ring swept-up by the fast WR wind to expand spherically despite 
                 of the driving star's huge bulk velocity. 
                 This simulation model has qualitative properties in accordance with 
                 the WR 124 and its surrounding nebular structure 
                 M1-67~\citep{sluys_aa_398_2003}. 
                 }                 
                 }      
        \label{fig:3}  
\end{figure}

\begin{figure}
        \centering
        \begin{minipage}[b]{ 0.485\textwidth} 
                \includegraphics[width=1.0\textwidth]{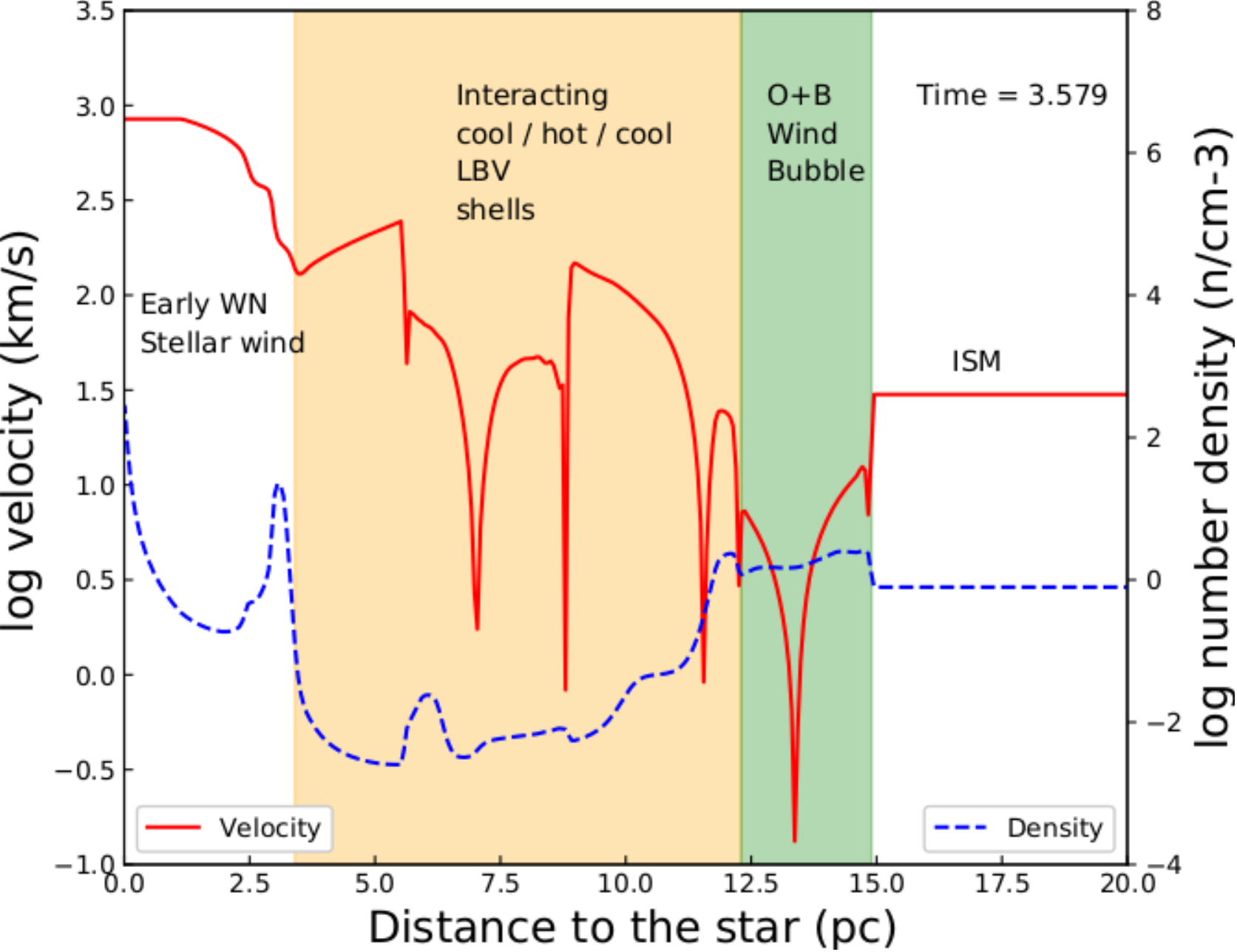}
        \end{minipage}     
        \caption{\mpo{Profiles of gas velocity (in $\rm km\ \rm s^{-1}$) and number density taken along the direction of 
             motion of the runaway star at time $3.579\, \rm Myr$ in our model Run-v30-n0.79, that involves} a star moving at 
             $30\, \rm km\, \rm s^{-1}$ (see also         
         Fig.~\ref{fig:cuts}).
            The orange and green regions highlight the regions mostly made of 
         LBV or O/B material. Note that additional mixing may happen.
                 }      
        \label{fig:plot_cuts_vel}  
\end{figure}

\textcolor{black}{
Fig.~\ref{fig:1}-\ref{fig:plot_cuts_vel} illustrate the results in our numerical models.
}
Fig.~\ref{fig:1} displays the gas number density (in $\rm cm^{-3}$) of our 
model Run-v30-n0.79 plotted in logarithmic scale (left-hand panels) 
and the corresponding \textit{Spitzer} $24$-$\mathrm{\mu m}$ infrared emission 
maps (right-hand panels). 
\mpo{To be noted from the figure is that the wind-ISM interaction generates a steady-state MHD bow-shock system composed} of 
an inner termination shock and an outer forward shock engulfing two regions 
of hot, low-density stellar wind and cold, dense shocked 
ISM~\citep{meyer_mnras_464_2017}. 
In Fig.~\ref{fig:1}b \mpo{one sees that the wind material of the B-type phase passes} through the 
main-sequence bow shock and produces \mpo{the appearance of} a Napoleon's Hat surrounding a 
cavity of low-density stellar wind. 

Density profiles \mpo{along 
the symmetry axis of model Run-v30-n0.79 taken at selected times are} shown in Fig.~\ref{fig:cuts}. 
The solid thin blue line illustrates the stellar surroundings 
at the WR time, with the expanding wind ($0$-$2\, \rm pc$), the 
WR shell ($2$-$2.5\, \rm pc$), the successive cold/hot/cold 
LBV phases engendering two shells ($5.5$-$7.5\, \rm pc$), 
the shock front of B spectral type that has reached and pushed outwards the 
former main-sequence stellar wind bubble ($10$-$12.5\, \rm pc$). 
From time $3.579\, \rm Myr$ the WR shell keeps on expanding 
(Fig.~\ref{fig:1}b,c), reach ($3.609\, \rm Myr$) and merges ($3.614\, \rm Myr$) 
with the shocked layers of O-type and B-type wind material. 
At later times (dotted black line), the WR material has merged with the 
main-sequence bow shock.

Fig.~\ref{fig:2} is as Fig.~\ref{fig:1} but for our model Run-v100-n0.79, i.e. a star with 
velocity $v_{\star}=100\, \rm km\, \rm s^{-1}$. The shaping of the circumstellar 
medium happens \mpo{in a similar way}, with a WR ring nebula developing inside the cavity of slow 
LBV gas (Fig.~\ref{fig:2}a). As an effect of the faster stellar 
motion, no Napoleon's hat forms and the ring interacts sooner with the bow shock 
since its stand-off distance is smaller~\citep{wilkin_459_apj_1996}. 
Such a ring does not behave like regular bow-shock-producing material, and a bright 
arc-like nebula \mpo{does not form, as is evident in} the infrared emission maps in Fig.~\ref{fig:1}-~\ref{fig:3},  
\textcolor{black}{
as a consequence of the dust spatial distribution. 
Fig.~\ref{fig:temperature} illustrates the location of the dust in the nebulae. 
It plots the gas density field (left, in $\rm g\, \rm cm^{-3}$) and the corresponding 
shocked material distribution (right, in $\rm K$). 
}
The portion of the ring that collides with the termination shock upstream of the 
stellar motion becomes denser, although its overall circular shape is conserved.

Fig.~\ref{fig:3} is \textcolor{black}{the same} as 
Fig.~\ref{fig:1} \textcolor{black}{but} for our model Run-v200-n0.01. The star 
moves \mpo{at very high speed} in a diluted medium taken to be $0.01\, \rm cm^{-3}$, 
which is typical of the low-density found in and above the Galactic plane.
Consequently, the stand-off distance of the pre-WR bow shock is 
huge, about $20\, \rm pc$, and so is the cavity of low-density stellar wind embedded inside of it. 
The shell of \textcolor{black}{the} fast WR wind expands into it and \mpo{assumes the shape of} a ring-like 
nebula. Due to the very extended shape of the bow shock, the ring has \mpo{lots of}
space and time to expand, \mpo{and consequently the} Rayleigh-Taylor instability develops. 
The curvature of the unstable WR ring is modified once it interacts 
with the contact discontinuity produced during the previous 
evolutionary phase; it adopts a \mpo{somewhat oblate morphology with higher density} upstream 
of the stellar motion. 
This mechanism naturally holds for high latitudes above the Galactic plane 
where the ISM density is dilute, which makes the bow shock fainter
and emphasizes the process of ring formation.

\textcolor{black}{
Note that the 2.5D nature of the MHD simulations implies a rotation of the 
solution around the $Oz$ axis of the computational domain when interfacing 
the MHD bow shock model with the radiative transfer code. Consequently, the 
projection of the Rayleigh-Taylor-unstable ring nebulae induces a series of 
stripe-like features in the infrared emission map (right-hand part of 
Figs.~\ref{fig:3}b,c), see also~\citet{meyer_obs_2016}. 
The detailed structure of the apex of MHD bow shocks might also be affected by 
the axis of symmetry as it can induce a concave-inward form that differs from 
the classical shape of stellar-wind bow shcoks, see Figs.~\ref{fig:3}b and 
Fig.~1 of~\citet{meyer_mnras_464_2017}. Similarly, artefacts can develop in 
the trail of bow shock nebulae when gas accumulates and pills-off along the 
symmetry axis, see Fig.~5 of~\citep{2020MNRAS.493.3548M}. 
However, the latter does not affect our results. Circumventing those 
2.5D-induced artefacts would require computationally-intensive full 3D 
MHD models. 
}

In Fig.~\ref{fig:plot_cuts_vel} \mpo{we show} the velocity profile along 
the direction of motion of the star at an early 
time of ring expansion in our model Run-v30-n0.79. 
The successive wind phases carve the main-sequence bow shock, 
making room for the \textit{fast} WR wind to expand into a 
\textit{slower} medium harboring multiple shocks and discontinuities 
of cold/hot LBV and earlier-type wind, hence producing 
a ring nebula. As the \mpo{speed of the slow gas} in the cavity is still larger than the 
stellar bulk motion, the ring seems to be co-moving with the runaway star.

\section{Discussion} 
\label{section:discussion}

In this paper, we presented numerical models for the circumstellar medium of very massive
runaway stars. 
\mpo{For a star with initial mass of $60\, \rm M_{\odot}$ stars that follows an 
evolutionary path successively characterised by phases with O and B spectral 
type, multiples cold/hot LBV eruptions,} and several 
final WR phases~\citep{groh_aa564_2014}, we investigate the formation of 
ring-like nebulae. 
Our simulations show that the stellar wind remains confined 
inside its main-sequence wind bubble. 
%
%
%
%
\mpo{We calculate the radiative transfer against 
dust opacity in the stellar environment and demonstrate that} the \mpo{ring-formation mechanism discussed above} explains the systematic centering 
of circumstellar shells surrounding many evolved massive stars such 
WR stars, that are observed as mid-infrared circular 
circumstellar nebulae~\citep{2010MNRAS.405.1047G}. 

\mpo{Previous models of $20$$-$$30\, \rm M_{\odot}$ runaway WR stars assumed a simpler evolution through a red-supergiant phase. They indicated} that the 
early WR wind expands spherically before eventually interacting
directly with the ISM, once the WR ring passed through the relatively compact red supergiant bow shock~\citep{brighenti_mnras_273_1995,vanmarle_aa_444_2005}.
%
The shells of WR wind get rapidly distorted once they interact with the ISM. 
For very massive stars, the rings develop in much larger cavities, 
and therefore the observation of a ring co-moving with very fast runaway 
WR stars is the consequence of a complex stellar evolution involving 
series of LBV eruptions. This constrains 
the star's initial mass to $>40\, \rm M_{\odot}$, i.e. to the mass range of 
stars which do not directly evolve to the WR stage via a \mpo{single} red 
supergiant phase.

%
%

Inside \textcolor{black}{the wind-blown} bubble, the post-main-sequence winds generate 
a pattern of various shocks and unstable discontinuities separating 
slow, dense, and cold gas from hot and fast material that collide with each other, 
creating a cavity of slow stellar wind inside of which the fast 
WR material is released. 
Star-centered rings such as the nebula surrounding WR71 are far denser and 
closer to the star than the bow shock, \mpo{and consequently they appear brighter in dust-scattered starlight} than the bow shock 
itself~\citep{Faherty_apj_147_2014,Flagey_aj_148_2014}. 
%
%
LBV winds are often anisotropic. Older rings may therefore lose their early sphericity \mpo{on account of propagation} in 
an anisotropic wind zone, which explains the variety of observed 
shells~\citep{2010MNRAS.405.1047G,stock_mnras_409_2010}. 

\mpo{For slower stars moving} in the Galactic plane, a Napoleon's hat 
\mpo{forms on account of} multiple arced structures arising from the succession of different 
phases (Fig.~\ref{fig:1}). 
Another trace of complex pre-WR evolution are the series of arcs
produced in the main-sequence bow shock when the various winds interact 
with the termination shock (Figs.~\ref{fig:2} and ~\ref{fig:3}). 
The aftermath of the succession of mass-loss events \mpo{is} visible 
at H$\alpha$ in the vicinity of WR~16's shell, see left panel 
of Fig.~1 in~\citet{toala_aa_559_2013}.

Our simulations compare well with high-latitude runaway WR stars. 
The properties of the early WN type star WR 124 with its 
nebular structure M1-67 and the star WR 148, both moving 
at velocities $v_{\star}\approx200\, \rm km\, \rm s^{-1}$, resemble 
our Run-v200-n0.001. 
WR 124 is not surrounded by a circular ring but a series of clumps, 
filaments and arcs, suggesting that the star just entered the WN 
phase and that no swept-up dust material has formed yet. \mpo{Associating the arcs to LBV 
ejections,}
\citet{sluys_aa_398_2003} derived the presence of a large-scale bow 
at $1.3\, \rm pc$ from the star. 
Such a small stand-off distance \mpo{is in conflict with} our model as we obtain 
$\approx 20\, \rm pc$, however using an ISM density of 
$n_{\rm ISM}\approx 0.01\, \rm cm^{-3}$ while the authors 
assumed $n_{\rm ISM}\approx 1.0\, \rm cm^{-3}$ which \mpo{we deem too large for a 
high-}latitude star. 
WR 124's bow shock is located at a larger distance and it is 
therefore much fainter compare to the nebular shell M1-67.

WR 148 has similar altitude and proper motion \mpo{as} WR124 but displays 
neither a bow shock nor a detected nebula. Our models \mpo{suggest} 
that the existing ring nebula is too young to be seen at such distance 
($>8\, \rm kpc$) and that a large, faint bow shock should surround it. 
Our model Run-v200-n0.01 assumes the highest known stellar velocity of a WR 
star, but \textcolor{black}{a very} dilute medium. Any star located higher 
above the Galactic disk will naturally produce an even larger cavity of 
slow stellar wind and rings of swept-up \textcolor{black}{hot} 
material~\citep{rate_mnras_493_2020}. 
This explains the systematic absence of observed bow shocks around 
high-latitude runaway WR stars. \mpo{They are} too large, diluted and faint to be 
detected, despite their huge proper motion and \textcolor{black}{significant}  
mass-loss rate~\citep{toala_aa_587_2015d}. 
\textcolor{black}{
Finally, our models explain the circumstellar structures around high 
high-latitude WR stars without assuming their in-situ formation.  
}

\textcolor{black}{
Last, the duration of the LBV phase might affect the development of the ring nebula. 
The stellar evolution model of~\citet{groh_aa564_2014} assume a rather long LBV 
phase of $2.35\times10^{5}\, \rm yr$, while much shorter 
LBV phases exist. e.g. in the case of lower-mass ($20$$-$$25\, \rm M_{\odot}$) 
progenitors~\citep{chita_aa_488_2008,groh_aa_550_2013} or at lower-metallicity 
such as in the Large Magellanic Cloud~\citep{bohannan_aspc_120_1997}. 
Shorter LBV eruptions will supply the interior of the bow shock with less  
material, and, consequently, diminish the amount of material accumulated in the ring 
that will become thinner as it expands outwards. This might affect our results, 
as long as the star is at rest or moves slowly~\citep{2020MNRAS.493.3548M}, whereas 
if it moves fast and/or in a dense medium the stand-off distance of the bow shock 
is short and less LBV material is required to generate a ring nebulae that will 
last less time. 
The length of the WR phase does not matter in the ring production as it acts as a source 
of momentum that swepts up the LBV material. 
All these elements are consistent with the conclusions of~\citet{Graf2012} who found 
that ring nebulae are mostly observed around early-type WR stars.
}

\textcolor{black}{
The models presented in this paper are computed of one particular  
stellar-evolution model~\citep{groh_aa564_2014}. The formation of ring 
nebulae around runaway WR stars should qualitatively 
remain the same regardless of the details of its evolution history, as 
long as the star undergoes episodes of faster and slower winds and ejects a
considerable amount of mass after its main-sequence phase (such as a LBV mass ejections). 
}


\section{Conclusion} 
\label{section:conclusion}

Our simulations change the paradigm of the surroundings of WR stars 
and reveal a complex picture the formation of \textcolor{black}{WR ring nebulae}. 
In the Galactic plane, the same fast wind-slow wind interaction mechanism \mpo{is}
responsible for the formation of observed near-infrared ring nebulae 
appearing to be co-moving with some runaway 
WR stars~\citep{2010MNRAS.405.1047G}, as long as a sufficiently 
complex evolution via, e.g., a B spectral-type phase and LBV 
eruptions happen and enlarge the wind cavity. The presence of infrared 
rings, \mpo{that are} brighter than the bow shock, therefore reflect a very \mpo{large}
($>40\, \rm M_{\odot}$) initial mass of these WR stars.

The absence of detected bow shocks and the presence of diffuse ring nebulae 
around high-Galactic-latitude WR stars does not 
imply the nonexistence 
of supersonic stellar motion though the ISM, \mpo{because that would  produce a distant and} very faint 
bow shock. 
%
%
%
Our study motivates further high-resolution, multi-wavelength observational 
campaigns of the circumstellar medium of Galactic WR stars such as 
the $3$-$\mathrm{mm}$ interferometric ALMA observations of Westerlund 
1~\citep{fenech_617_aa_2018} or the \mpo{observations of} non-thermal radio emission of WR 
wind bubble G2.4+1.4~\citep{prajapati_apj_884_2019}, in order to unveil their 
detailed structures and to constrain the late evolution of their driving star.


\section*{Acknowledgements}

The authors acknowledge the North-German Supercomputing Alliance 
(HLRN) and the Max Planck Computing and Data Facility (MPCDF) 
for providing HPC resources. 
\textcolor{black}{
LMO acknowledges financial support by the Deutsches Zentrum f\"ur Luft und 
Raumfahrt (DLR) grant FKZ 50 OR 1809, and partial support by the Russian 
Government Program of Competitive Growth of Kazan Federal University. 
}
     

\bibliographystyle{mn2e}

\footnotesize{
\bibliography{grid}
}


\end{document}